# Are Advanced Potentials Anomalous?

Michael Ibison

*Institute for Advanced Studies at Austin*
*4030 West Braker Lane, suite 300, Austin, Texas 78759, USA.*

**Abstract.** Advanced electromagnetic potentials are indigenous to the classical Maxwell theory. Generally however they are deemed undesirable and are forcibly excluded, destroying the theory's inherent time-symmetry. We investigate the reason for this, pointing out that it is not necessary and in some cases is counter-productive. We then focus on the direct-action theory in which the advanced and retarded contributions are present symmetrically, with no opportunity supplement the particular integral solution of the wave equation with an arbitrary complementary function. One then requires a plausible explanation for the observed broken symmetry that, commonly, is understood cannot be met by the Wheeler-Feynman mechanism because the necessary boundary condition cannot be satisfied in acceptable cosmologies. We take this opportunity to argue that the boundary condition is already met by all expanding cosmologies simply as a result of cosmological red-shift. A consequence is that the cosmological and thermodynamic arrows of time can be equated, the direct action version of EM is preferred, and that advanced potentials are ubiquitous.

## MAXWELL THEORY

### Inherent time-symmetry of the theory

In the Lorentz Gauge $\partial \circ A = 0$ the Maxwell 4-potential $A$ generated by a four current $j$ satisfies the wave equation $\partial^2 A = j$, where the symbol $\circ$ denotes the Lorentz scalar product. A completely general solution can be written in terms of Green's function

$$A = G * j \tag{1}$$

where the star denotes convolution. The generality, despite the lack of an explicit complementary function, is due to the indeterminacy of $G$ up to a solution of the homogeneous wave equation. That indeterminacy can be removed by imposition of an appropriate boundary condition on $A$.

Use of the *retarded* Green's function

$$G(x) = G_{ret}(x) = \frac{\delta(t - |\mathbf{x}|)}{4\pi |\mathbf{x}|} \tag{2}$$

to compute a retarded potential is a special case, consistent with the boundary condition that the potential vanishes at all times *before* the current is activated (observe that $G_{ret}(x) = 0$ when $t < 0$). In the terminology introduced below, this is the 'Sommerfeld radiation condition' that $A_{in} = 0$. An idealized example is an initially quiescent system in which a mechanically driven current oscillator is switched on at $t = 0$. In that case one could use (2) in (1) to compute the future potentials for all $t > 0$ from the motion of the current during $t > 0$, and one can speak unambiguously of *emission of radiation* from the oscillator ('*source*'). Alternatively one could refer to the same process as the *production of retarded radiation* in order to highlight the quality that the phase-fronts of the potential diverge from a point source in (forwards) time.

Use of the *advanced* Green's function

$$G(x) = G_{adv}(x) = \frac{\delta(t + |\mathbf{x}|)}{4\pi |\mathbf{x}|} \tag{3}$$

to compute an advanced potential is a special case, consistent with the boundary condition that the potential is zero everywhere at all times *after* the current has become deactivated (observe that $G_{adv}(x) = 0$ when $t > 0$). In the terminology introduced below, this is the condition that $A_{out} = 0$. An idealized example is an initially active system with non-zero potentials in which a mechanically damped current oscillator finally becomes quiescent at $t = 0$. In that case one could use (3) in (1) to compute the historical potentials for all $t < 0$ from the motion of the current during $t < 0$, and one could speak unambiguously of *absorption of radiation* by the oscillator. In the case of absorption, vanishing of the potentials is a *future* boundary condition. The domain of the time coordinate of the fields *and* the current is the same, just as it is in the case of emission. The causal order linking potential and current however, is the reverse of that in the purely retarded case.

> Comparison of these two cases gives that the absorption of retarded radiation can equally be regarded as the *production of advanced radiation*. In any case, advanced potentials are a necessary component of a Maxwell theory formulated as (1).

## Asymmetry of the formalism

The unambiguous designation of potentials as advanced or retarded relies on the special boundary conditions as described above and also on the restriction that the solution of the inhomogeneous wave equation be written exclusively as a particular integral. But an alternative is to write the solution exclusively in terms of the retarded Green's function, and satisfy the boundary conditions with the aid of a complementary function '$A_{in}$':

$$A = G_{ret} * j + A_{in}; \quad \partial^2 A_{in} = 0. \tag{4}$$

If the current is switched on at $t = 0$, then $G_{ret} * j$ vanishes for $t < 0$, and therefore $A_{in}$ is the 'incoming' field present at earlier times. We can use (4) even for the special case of purely advanced potentials due to perfect absorption described above:

$$A = G_{adv} * j. \tag{5}$$

Comparison of (4) with (5) gives

$$A_{in} = -2G_- * j \tag{6}$$

where

$$G_-(x) \equiv \frac{1}{2}\left(G_{ret}(x) - G_{adv}(x)\right) = \frac{1}{2}\left(\frac{\delta(t-|\mathbf{x}|)}{4\pi|\mathbf{x}|} - \frac{\delta(t+|\mathbf{x}|)}{4\pi|\mathbf{x}|}\right). \tag{7}$$

It is easily verified that $G_-$ is a solution of the (homogeneous) wave equation, and so, therefore, is (6). The interpretation now is that the complementary function given by (6) is a 'pre-existing state' of the vacuum. Onto this vacuum mode is superposed the retarded potential $G_{ret} * j$. The retarded parts cancel and only the advanced part remains.

## Reasons for the broken symmetry

Granted an arbitrary complementary function it is clear from the above that through use of (4) one can always choose to model the effect of a current as producing an exclusively retarded potential, regardless of how much absorption or emission is taking place. This seems to be the commonly accepted use (or interpretation) of the classical Maxwell theory. We emphasize however that once an arbitrary complementary function is admitted the attribution of *retardation* to the potentials no longer has a definite meaning; it implies a distinction (between advanced and retarded) without a (physical) difference.

There are two - possibly related reasons – for the common adoption of this asymmetry, despite the intrinsic symmetry of the theory. Firstly, it is a fact of observation that in our particular universe momentum flux of radiation more commonly leaves - rather than enters - a closed volume. Note that there is no *intrinsic* time-asymmetry in the physics of electromagnetism that could explain this fact; the same mathematics adequately describes a movie of purely electromagnetic interactions run forwards *or* backwards in time.

In a closed purely electromagnetic universe this bias in favor of flux divergence away from point sources could, for example, be attributed to an initial condition of disequilibrium between matter and EM field, with the matter being, on average, hotter than the field[1]. The diverging flux can then be interpreted as the signature of a movement toward equilibrium with increasing entropy. Consistent with the discussions above however, the fact of the observed asymmetry of our particular universe does not mandate the exclusive use of (4), since, even in this case, one could just as well use

$$A = G_{adv} * j + A_{out}; \quad \partial^2 A_{out} = 0. \tag{8}$$

There is, however, a computational advantage to the use of a particular form: if the boundary condition on the potentials is given say as $A, \partial A / \partial t = 0$ at $t = 0$, then the complementary function will be zero more often (over a random sample of matter) if one uses (4) rather than (8) or (5), in which case the Sommerfeld radiation condition is met, and

$$A = G_{ret} * j. \tag{9}$$

A second reason for the choice (4) is the time-asymmetric disposition the observer brings to the observation - an issue briefly discussed below under the heading 'Anthropocentrism'.

## Exclusively Advanced Potentials

It is illustrative to contrast the relatively familiar behavior of emission requiring exclusively retarded potentials with typical electromagnetic interactions in an imaginary anti-Boltzmann universe, characterized by an initial condition in which the field is hotter than the matter[2]. Radiation flux in the anti-Boltzmann universe would more often converge upon - rather than diverge from - matter. It would not be uncommon in the anti-Boltzmann universe to find matter that is colder than the background radiation, just as it is not uncommon to find matter that is hotter than the background in our universe. Likewise, anti-Boltzmann matter that is hotter than the background would require special preparations, as are required to cool matter below 2.7 K in our universe. Whereas in our Boltzmann universe it is often a good approximation to ignore whatever fields are present at $t = 0$, and compute the future potentials using (9), in the anti-Boltzmann universe the default disequilibrium condition is that charged matter is colder than the background and destined to heat up in order to come to equilibrium[3]. The (yet-to-be-attained) equilibrium current and potential will satisfy $\partial^2 A_{equ} = j_{equ}$. The departures $\Delta A \equiv A - A_{equ}$, $\Delta j \equiv j - j_{equ}$ satisfy

$$\partial^2 \Delta A = \Delta j; \quad \Delta A = 0 \ \forall \ t > 0$$
$$\Rightarrow \Delta A = G_{adv} * \Delta j \tag{10}$$

confirming that the common electromagnetic interactions in the anti-Boltzmann universe involve advanced rather than retarded potentials. Here $t = 0$ is the time at which the oscillator has absorbed sufficiently from the background that it has come to complete equilibrium.

---

[1] We postpone consideration of the effect of cosmological expansion.
[2] Here the term 'anti-Boltzmann' refers only to the evolution of matter – which tends to *absorb* energy. The anti-Boltzmann universe still evolves according to the second law of thermodynamics - entropy increases as the matter gets hotter and the fields colder. Consequently an anti-Boltzmann universe would *not* look like a time-reversed image of our universe because the time development of the two universes do not cross over but meet – statistically - at the equilibrium point, approached from two different directions.

We observe that the primary quantities of interest for the anti-Boltzmann physicists will be the departures from the background, just as our primary quantities are departures from the zero-point states (or perhaps zero-point plus Cosmic Microwave Background states) of matter and field. Accordingly, under special conditions those background fields can be used to simulate in our universe the common electromagnetic exchanges predicted of the anti-Boltzmann universe. An example is a Casimir cavity of appropriate geometry[4] that is permitted to collapse adiabatically. In this case the final 'out' field is the ZPF with the cavity gone. Supposing that the potentials can be modeled classically (e.g. [2]) then that out field is the 'true' ZPF – i.e. the ZPF in a universe with no cavity. Then

$$\Delta A = G_{adv} * j; \quad \Delta A = A - A_{ZPF}, \ \partial^2 A_{ZPF} = 0, \ j, \Delta A = 0 \ \forall \ t > 0, \tag{11}$$

where $j$ contains all the macroscopic currents in the cavity walls that respond to the EM fields. $t = 0$ is the final state of affairs with collapsed cavity and perfectly restored ZPF. (Strictly, this can only be at $t = \infty$ of course, but we can assume a finite time if we are prepared to restrict our vision to the local environment around the cavity.) From the perspective of the complete set of vacuum modes of the final ZPF, the 'initial' state of affairs with cavity plates separated is a non-ZPF state with special phasing of the vacuum modes conspiring to reduce the field inside the cavity.

---

[3] This is a static (not-expanding) universe. In order to have a finite amount of energy we must ignore the ZPF or else have a cutoff. In a classical version we must assume a cutoff for the Rayleigh-Jeans spectrum.

[4] A pair of plates experiences an inward pressure, but in a conducting sphere the pressure is radially outwards [1].

## Summary

We now summarize the major points above:

i) There is an unambiguous decomposition of the potentials into 'advanced' and 'retarded' only if no complementary function is admitted in the solution of the wave equation for the potentials.

ii) If a complementary function *is* admitted, then one is completely free to designate the particular integral as comprised of any combination of advanced and retarded potentials that one chooses.

iii) The traditional Maxwell approach is to exercise the option of admitting a complementary function and then express the particular integral exclusively in terms of the retarded Green's function. This choice, however, does not impact, i.e. constrain, the physics.

## DIRECT ACTION

### Relation to the Maxwell theory

Initially we write the total action as $I = I_m + I_{EM}$, where $I_m$ is the matter action

$$I_m = -\sum_i m_i \int dt \sqrt{1 - \mathbf{v}_i^2} \; , \qquad (12)$$

and $I_{EM}$ is the electromagnetic action which can be written

$$I_{EM} = \int d^4x \left( \frac{1}{2} A \circ \partial^2 A - j \circ A \right) \qquad (13)$$

where $j$ is the total 4 current of all the sources[5]. As required, variation of $A$ gives $\partial^2 A = j$. In the direct action version of EM however, the potential is not an independent degree of freedom and so can be eliminated from the action. We choose

$$I_{EM} = -\frac{1}{2} \int d^4x \int d^4x' \delta\left((x-x')^2\right) j(x) \circ j(x'), \qquad (14)$$

which corresponds to the substitution into (13)

$$A = G_+ * j \qquad (15)$$

---

[5] Despite appearances the action remains gauge invariant due to charge conservation – see [2].

where

$$G_+(x) \equiv \frac{1}{2}\left(G_{ret}(x) + G_{adv}(x)\right)$$
$$= \frac{1}{2}\left(\frac{\delta(t-|\mathbf{x}|)}{4\pi|\mathbf{x}|} + \frac{\delta(t+|\mathbf{x}|)}{4\pi|\mathbf{x}|}\right) = \frac{1}{4\pi}\delta(x^2) \quad (16)$$

is the time-symmetric Green's function. If the current is written in terms of contributions from individual charges, $j = \sum_k j_k$, one has [3-5]

$$I_{EM} = -\frac{1}{2}\sum_{k,l} \int d^4x \int d^4x' \delta\left((x-x')^2\right) j_k(x) \circ j_l(x'). \quad (17)$$

Given the elimination of the potentials in the classical theory (and, even more impressively, an infinite dimensional Hilbert space in the quantized theory), a successful direct-action theory (i.e. consistent with observation) would be infinitely more parsimonious from an Ockham's razor perspective than the Maxwell theory. However, given that the action (17) describes currents that interact directly with each other across space and time rather than through the intermediate *local* action of potentials, the loss of locality is a source of discomfort for many. Using (15) however, the potentials *can* be re-introduced, and Lorentz forces etc. computed in the usual way. But in that case - unlike in the Maxwell theory - there is no choice in the Green's function - the time-symmetric Green's function is mandatory. And there is no complementary function - there are no 'in' or 'out' fields. Consequently one can speak unambiguously of advanced and retarded potentials generated by the currents.

## History

An early attraction of the direct action theory was the possibility of explicit exclusion of the troublesome electromagnetic self-action terms, achieved simply by removing the diagonal terms $k = j$ in the double sum[6]. Later it was realized that self-action must be present in the quantized matter version of the theory, since particle creation is a form of promoted self-action [7]. A related argument can be given for the classical theory [8]. And self-action was found to be necessary in the subsequent development of a relativistic quantized-matter version of the direct action theory [9-12]. Further, a recent development of the classical version of the direct-action theory obtains some interesting results that rely crucially on the *retention* of self-action [13].

In 1945 [14] and 1949 [15] Wheeler and Feynman offered an explanation within the (intrinsically time-symmetric) direct action theory for the observed time-asymmetry of our universe. Briefly, they showed that despite the appearance of radiation by matter into the vacuum, that if in fact all such energy is destined to be absorbed then it can be regarded as a direct current-current interaction and therefore consistent with the direct action theory. Whereas the analysis of Wheeler and Feynman took place in the framework of a static universe, subsequently Davies [16,17] showed that in none of the various plausible cosmologies is there appreciable absorption; radiated emitted from our era is almost certain to escape to infinity during the expansion. Failure to meet the Wheeler and Feynman boundary conditions apparently falsified the direct-action theory. The reader is referred to Pegg [18] for a more comprehensive historical review of the direct-action theory.

---

[6] Leiter [6] has shown how the same technique can be applied to the Maxwell theory by keeping track of the fields associated with each current source.

## THE COSMOLOGICAL BOUNDARY CONDITION

### Cosmological red-shift

The Wheeler Feynman boundary condition of the perfect future absorber demanded that all EM radiation energy end up in the material degrees of freedom. We show here however that it is sufficient for the direct action theory that the EM radiation energy goes to zero at infinity, regardless of the mechanism.

The Friedmann equation for the FRW metrics is

$$\left(\frac{1}{a}\frac{da}{dt}\right)^2 = \frac{8\pi G}{3}\rho(a) - \frac{k}{a^2} \qquad (18)$$

where

$$\rho(a) = \rho_\Lambda + \frac{\rho_{m0}}{a^3} + \frac{\rho_{r0}}{a^4} \qquad (19)$$

explicates the various contributions to the energy density. Since the proper volume goes as the coordinate volume times $a^3$, the total proper energy density of matter is constant throughout the expansion, whilst the total proper energy density of radiation falls like $1/a$. Radiation energy is not conserved! (See [19], but in particular Baryshev [20] for an interesting discussion of this issue.) This is rather surprising since the derivation of (18) is contingent on the decoupling of matter and radiation. Where does the energy go? One might think that it is lost in overcoming the gravitational field, just as light radiated by a massive body suffers a gravitational red-shift. But this explanation works only if it can be shown that the gravitational field is correspondingly augmented – i.e. that energy is conserved. The problem is that GR does not provide a mechanism to compute the gravitational field energy as a contribution to a conserved total.

Let us write (18) in the suggestive form

$$\frac{1}{2}\left(\frac{da}{dt}\right)^2 = \frac{4\pi G}{3}\left(a^2\rho_{\Lambda 0} + \frac{\rho_{m0}}{a} + \frac{\rho_{r0}}{a^2}\right) - k/2 \qquad (20)$$

and recall the work of Harrison, who has shown that the Friedmann equation with $\rho_\Lambda = \rho_r = 0$ has the surprisingly straight-forward Newtonian interpretation that $a(t)/a(0)$ is the (normalized) radius to a 'co-moving' surface in *Euclidian 3-space* of a non-relativistic explosion (or implosion) of matter bound by the collective gravitational field. No appeal to the hyper-surface and embedding space used by Misner, Thorne and Wheeler [21] is necessary. The meaning of the terms in Newtonian interpretation is quite different from those in the traditional FRW / GR interpretation. One interprets the density terms on the right hand side of (20) as components of a negative gravitational potential, the constant *k* as (proportional to) the total energy, and the left hand side as the total kinetic energy of the expansion. An advantage of going over to this picture therefore is that one can speak unambiguously of *gravitational energy*. Let us borrow from the work of Harrison and include now the radiation and vacuum terms in the same picture. By differentiation of (20)

$$\frac{d^2 a}{dt^2} = \frac{4\pi G}{3}\left(2a\rho_{\Lambda 0} - \frac{\rho_{m0}}{a^2} - 2\frac{\rho_{r0}}{a^3}\right) \qquad (21)$$

we learn that the radiation is gravitationally bound with a force that goes as $a^{-3}$ rather than $a^{-2}$. Though Newton may be surprised by the peculiar distance dependency of the gravitational binding force of the radiation, he would see that the force is conservative (since it can be expressed as a function of distance only) and would not doubt that energy was being conserved: the work done in overcoming the binding force is exactly the loss of kinetic energy in the expansion, regardless of the relative contributions of matter and radiation.

## Lossless propagation of EM fields

The Newtonian-inspired relationship between the radiation density and expansion rate is true of course in the FRW / GR equations, though one is not entitled there to refer to the gravitational binding energy of radiation, or to identify to the expansion rate (squared) as a kinetic energy. The more traditional point of view rests to some degree on a particular method of analysis, exploiting the conformal invariance of Maxwell equations, applicable here because the three Friedmann-Robinson-Walker metrics can be written in conformal coordinates. Here we will consider only the flat space metric for which *k* = 0 in (18) for which the line element is commonly written as

$$ds^2 = dt^2 - a^2(t)d\mathbf{x}^2. \qquad (22)$$

The conformal form is obtained by the coordinate change $ad\eta = dt$ whereupon

$$ds^2 = a^2\left(d\eta^2 - d\mathbf{x}^2\right). \qquad (23)$$

In conformal coordinates the Maxwell equations retain their Minkowski form. In particular, where there are no sources one still has $\partial^2 A_\mu = 0$ where $\partial^2$ is the flat-space (conformal time) D'Alembertian, and

$$F_{\mu\nu} = \partial_\mu A_\nu - \partial_\nu A_\mu. \tag{24}$$

The energy density is

$$\begin{aligned}\rho_r = T^0{}_0 &= -F_{0\nu}F^{0\nu} + \frac{1}{4}F_{\mu\nu}F^{\mu\nu} \\ &= -F_{0\nu}F_{\lambda\kappa}g^{\lambda 0}g^{\kappa\nu} + \frac{1}{4}F_{\mu\nu}F_{\kappa\lambda}g^{\mu\kappa}g^{\nu\lambda}\end{aligned}. \tag{25}$$

Using that the conformal metric is $g_{bc} = \eta_{bc}a^2$ where $\{\eta_{bc}\}$ is the Minkowski metric, the above is

$$T^0{}_0 = \left(-F_{0\nu}F_{\lambda\kappa}\eta^{\lambda 0}\eta^{\kappa\nu} + \frac{1}{4}F_{\mu\nu}F_{\kappa\lambda}\eta^{\mu\kappa}\eta^{\nu\lambda}\right)/a^4. \tag{26}$$

Let us refer to the entries in the covariant tensor $F_{\mu\nu}$ by **E** and **B** in the usual way,

$$\begin{aligned}\mathbf{E} &\equiv \{E_i\}; \quad E_i \equiv -\partial_0 A_i - \partial_i A_0 \\ \mathbf{H} &\equiv \{H_i\}; \quad H_i \equiv (\nabla \times \mathbf{A})_i\end{aligned}. \tag{27}$$

It is easily seen that

$$\partial^2 \mathbf{E} = 0, \quad \partial^2 \mathbf{H} = 0; \tag{28}$$

the fields propagate as if in Minkowski space-time. (Recall that we are working in conformal coordinates here, so that $\partial_0$ is differentiation with respect to the conformal time.) With (27), (26) becomes

$$\rho_r = T^0{}_0 = \frac{1}{2a^4}\left(\mathbf{E}^2 + \mathbf{H}^2\right) \tag{29}$$

showing the expected $1/a^4$ fall off with expansion of the coordinate density of the radiation energy. The proper energy density therefore falls as $1/a$

$$\sqrt{-\gamma}T^0{}_0 = \frac{1}{2a}\left(\mathbf{E}^2 + \mathbf{H}^2\right) \tag{30}$$

where $\gamma$ is the determinant of the spatial part of the metric tensor, $\gamma = \det\{g_{ij}\}$, where $i$ and $j$ index the spatial coordinates. Eq. (30) demonstrates the well known result that the energy of EM radiation in a proper volume is not conserved.

## Lossy propagation

A problem with this presentation is that the energy density is expressed asymmetrically in terms of the elements **E** and **H** of a *covariant* tensor. As a result, these fields do not appear in the final expression for $T^0{}_0$ - which is mixed - in the same way that do in Minkowski space. To be specific, the energy loss suffered by radiation during propagation is hidden through use of **E** and **H** defined in (27), because these fields satisfy (28) and therefore propagate losslessly. The energy loss appears in the expression (29) only as a result of the metric. This shortcoming can be overcome by associating fields with the elements of the mixed tensor $\{F^a{}_b\}$:

$$\tilde{\mathbf{E}} = \mathbf{E}/a^2, \quad \tilde{\mathbf{H}} = \mathbf{H}/a^2, \tag{31}$$

where **E** and **H** remain as defined in (27). With these the (coordinate) radiation energy density has the more familiar form

$$\rho_r = T^0{}_0 = \frac{1}{2}\left(\tilde{\mathbf{E}}^2 + \tilde{\mathbf{H}}^2\right). \tag{32}$$

The new fields propagate according to

$$\partial^2(\mathbf{E},\mathbf{H}) = 0 \Rightarrow \partial^2\left(a^2\left(\tilde{\mathbf{E}},\tilde{\mathbf{H}}\right)\right) = 0. \tag{33}$$

Expanding the D'Alembertian one obtains

$$\frac{\partial^2 \tilde{\mathbf{E}}}{\partial \eta^2} - \nabla^2 \tilde{\mathbf{E}} + \frac{2}{a^2}\frac{da^2}{d\eta}\frac{\partial \tilde{\mathbf{E}}}{\partial \eta} + \frac{1}{a^2}\frac{d^2 a^2}{d\eta^2}\tilde{\mathbf{E}} = 0. \tag{34}$$

The term proportional to $a^{-2}d^2a^2/d\eta^2$ has the character of an effective mass. The field loses energy with time as a consequence of the term proportional to $\partial \tilde{\mathbf{E}}/\partial \eta$. Note however that although the coordinate energy density falls as $1/a^4$, only $1/a$ of this factor is attributable to an effective cosmological damping, namely the loss of energy in a fixed proper volume associated with cosmological red-shift. The remaining factor of $1/a^3$ is the ratio of coordinate to proper volume, and is the same geometric factor as that suffered by matter. Let us demonstrate these effects for the case of an exponential expansion in conformal time

$$a = \exp(H_0\eta), \qquad (35)$$

which is a linear expansion in ordinary (FRW) time,

$$t = \int d\eta\, a = (\exp(H_0\eta)-1)/H_0 \Rightarrow a = 1 + H_0 t. \qquad (36)$$

Putting (35) in (34) one obtains

$$\frac{\partial^2 \tilde{\mathbf{E}}}{\partial \eta^2} - \nabla^2 \tilde{\mathbf{E}} + 4H_0 \frac{\partial \tilde{\mathbf{E}}}{\partial \eta} + 4H_0^2 \tilde{\mathbf{E}} = 0. \qquad (37)$$

For a single plane wave $\tilde{\mathbf{E}} \sim \exp(i\omega\eta - i\mathbf{k}\cdot\mathbf{x})$ one has

$$-\omega^2 + k^2 + 4i\omega H_0 + 4H_0^2 = 0. \qquad (38)$$

For any reasonable frequency $H_0 \ll \omega$ whereupon the effective mass correction can be ignored and

$$\omega \approx k + 2iH_0. \qquad (39)$$

Consequently one can account for the total energy loss (due to cosmological damping and the geometric factor) as due to plane wave decay by a factor of (approximately) $\exp(-2H_0\eta)$ - very nearly independent of frequency[7]. The effect on the electric field of the propagation loss is the same as the effect of absorption by matter; they both appear as a damping term in the wave-equation. Since *all* EM energy is dissipated during propagation, the future boundary condition provided by cosmological red-shift is precisely that sought by Wheeler and Feynman in order to justify the direct-action version of EM.

## Analogue Gravity

This above point of view effectively rests on the re-interpretation of the components of the gravitational metric as ordinary fields in Minkowski space through the substitution $g_{bc} = \eta_{bc} a^2$, whereupon a mixed GR tensor $T$ is replaced by a mixed Lorentz tensor $\tilde{T}$ according to

$$\begin{aligned}
T_{p_1 p_2 \ldots p_m}{}^{q_1 q_2 \ldots q_n} &= g^{r_1 q_1} g^{r_2 q_2} \ldots g^{r_n q_n} T_{p_1 p_2 \ldots p_m r_1 r_2 \ldots r_n} \\
&= a^{-2n} \eta^{r_1 q_1} \eta^{r_2 q_2} \ldots \eta^{r_n q_n} T_{p_1 p_2 \ldots p_m r_1 r_2 \ldots r_n} \\
&= a^{-2n} \tilde{T}_{p_1 p_2 \ldots p_m}{}^{q_1 q_2 \ldots q_n}
\end{aligned} \quad (40)$$

---

[7] Instead of (31) one could define fields $\hat{\mathbf{E}} = \mathbf{E}/a^{1/2}$, $\hat{\mathbf{H}} = \mathbf{H}/a^{1/2}$ for which the *proper* energy density has the Minkowski form: $\sqrt{-\gamma}\rho_r = (\hat{\mathbf{E}}^2 + \hat{\mathbf{H}}^2)/2$. The wave equation becomes $\partial^2 \hat{\mathbf{E}}/\partial\eta^2 - \nabla^2 \hat{\mathbf{E}} + H_0 \partial\hat{\mathbf{E}}/\partial\eta + H_0^2 \hat{\mathbf{E}}/4 = 0$, and the decay term is now exclusively attributable to cosmological damping - with no contribution from the geometric factor. However, the results will not differ qualitatively from those obtained using (31).

One then 'forgets' the metric origins of the scale factor, which is now interpreted as a field in flat space-time. And rather than following geodesics in curved space-time, the free EM field is interpreted as coupled to, and therefore *interacting with*, the scalar field $a(t)$ in Minkowski space-time. Clearly there is no problem applying this interpretation to the equations of motion once they have been obtained from variation of the $g_{\mu\nu}$ and the EM potentials $A_\mu$. Indeed, Landau and Lifshitz [22] pose as an exercise the reformulation of an *arbitrarily* curved space-time vacuum as a polarizable medium in Minkowski space-time. (The assignments (31) turn out to be a special case.) We postpone to a future effort derivation of the equations of motion for $a$ and $A$ from an action with just those - and no other - degrees of freedom. This will permit identification of a canonical Hamiltonian for the scale factor, and therefore an unambiguous definition of *conserved energy* in the Cosmological gravitational field. Of course one must still end up with (20). But the advantage will be a more formal justification for the explanation for that loss of energy of radiation is due to *interaction* with a gravitational field.

## Interpretation

In the direct action theory the EM fields of a single source are not exclusively retarded but are time-symmetric. The appearance of pure retardation is now explained as the result of interference by time-symmetric exchanges with the cosmological gravitational field. Just as the effect of a dielectric continuum can be regarded as the final result of a series of absorptions and re-emissions on the microscopic level [23], the macroscopic exchanges with the gravitational field implied by (34) can be interpreted likewise. If each exchange is subject to the constraint that it be time-symmetric, then the gravitational damping plays the same role as do the future absorbers in the Wheeler-Feynman theory. Anti-phase advanced waves from these exchanges arrive back at the current source to re-enforce the retarded component and cancel the advanced component. Consequently these proposed cosmological boundary conditions guarantee that every 'photon' (of which, strictly, there are now none) will be absorbed. The absorption is not by matter, but the whole system - which includes a term for the energy in the cosmological gravitational field. A 'prediction' of this implementation of the cosmological boundary condition is that, if the universe were not expanding, then there would be no apparent predominance of retarded radiation. Consequently the future state of the universe is felt in the present. If these arguments stand then the direct action theory is validated (and therefore preferred), and advanced potentials in the sense of (15) and (16) are ubiquitous.

# TIME-ASYMMETRY

## The arrows of time

Let us first of all identify an arrowless 'passive coordinate time' as a tool with which to parse the meaning of the various other times. This time is a continuous passive index into a 'pre-existing' time dimension of Minkowski space-time. Below we will have cause to doubt - in the event that space-time is curved - that the passive coordinate time is the same as the GR time coordinate.

A fact of observation is a predominance of radiation of EM fields over absorption *by matter*, which we can to refer to as an electromagnetic arrow of time. We wish to identify this with a locally defined thermodynamic arrow of time. It would be desirable to have a detailed accounting of the evolution of entropy with respect to coordinate time in a direct-action universe. Given the closure of the direct-action universe however, without further investigation it is not clear (at least to the author) that the total system entropy must be increasing. Even so, there is no doubt that the entropy as defined thermodynamically increases with respect to coordinate time *locally* – disregarding any entropy changes attributable to the cosmological gravitational field. Further, if one accepts the characterization of the cosmological boundary condition given above, then the local electromagnetic and therefore thermodynamic arrows of time are tied to the cosmological arrow of time. In particular, it is doubtful there could be an evolution of thermodynamic time in a static direct-action universe. Finally therefore, in a direct action universe justified by the boundary condition given above one has

$$t^{\uparrow}_{EM} = t^{\uparrow}_{\substack{local \\ thermodynamic}} = t^{\uparrow}_{\substack{local \\ entropic}} = t^{\uparrow}_{cosmological}. \qquad (41)$$

In the above we mean to imply that the cosmological time is a monotonic function of the scale factor. Traditionally the GR coordinate time is taken to be the same passive reversible index as appears in the Dirac equation. It is not clear however that these two times really are the same. Let us consider the implications that follow from forcing GR to refer instead to these arrowed times by eliminating altogether any reference to a symbol *t* in favor of the cosmological scale factor. We accomplish this by writing the FRW metric (22) as

$$ds^2 = da^2 / \dot{a}^2 - a^2 d\mathbf{x}^2 \qquad (42)$$

and substitute from the first Friedmann equation (18) to eliminate the time derivative. The cosmological line element then becomes[8]

---

[8] It is interesting that the scale factor becomes the conformal time in the case that $k = 0$ and the expansion is radiation-dominated.

$$ds^2 = \frac{da^2}{\frac{8\pi G}{3}a^2\rho(a)-k} - a^2 d\mathbf{x}^2. \tag{43}$$

The metric corresponding to this line element causes the Einstein tensor $G_{\mu\nu}$ to be *identically equal* to the cosmological stress energy tensor, i.e. Einstein's equation becomes an identity:

$$G_{\mu\nu} = R_{\mu\nu} - \frac{1}{2}Rg_{\mu\nu} \equiv 8\pi G T_{\mu\nu}. \tag{44}$$

Cosmological collapse cannot now be described without making a coordinate transformation that is not isomorphic with *a*, the implication being that the arrowed times in (41) and therefore practical clocks would stop (incrementing) if the universe stopped expanding.

By contrast, the time parameter of the Dirac and direct-action equations of motion is *reversible*, and so should be identified with a passive reversible index. Since quantum mechanical measurement processes are not time-reversible, it seems natural that they be indexed with the cosmological time. Further, since QM measurement apparently requires interaction with a 'classical' measuring instrument, it is reasonable to make this correspondence for the evolution of all classical systems, whether or not they are explicitly involved in a quantum-mechanical measurement. If so, then one can identify all classical times with the cosmological time, of which (43) is therefore just one example. And since GR is a classical theory, this position provides a (retrospective) justification for elimination of the passive time coordinate from the metric that led to (43).

The above reasoning invites the speculation that the discontinuity resulting from a measurement performed on a QM system arises because the mathematics of QM only correctly describes a system that is isolated from the cosmological expansion – specifically: the (alleged) cosmological gravitational field.

## Anthropocentrism

In considering the issue of reverse causation we should be alert to the possibility of anthropocentric bias for which there may be no physical justification. One problem for example is that the common subjective view of time is not explained by the mathematics. From the standpoint of the four-dimensional space-time block universe, the fields and particles can be regarded as 'pre-existing', the latter as one-dimensional world lines. By contrast, comparing the physics to the experience of the observer, it appears that the observer experiences successive adjacent slices through space-time, always in the direction from past to future. In other words, the observer experiences a unidirectional moving 'now'[9,10]. From the perspective of the moving now the one-dimensional particle world lines then appear as zero-dimensional points moving through space. That is, the particles acquire the quality of 'movement through time' from the observer, not the other way around. We agree with Squires [24] that the origin of this 'now' is deep mystery, not explained by current physics. Does our experience of a moving now require a revision of our physics?

---

[9] There is some danger in speaking of 'movement through time', since the notion of movement is grounded in the perception of movement through space, which in this context is logically anterior to movement through time.

More readily dismissed is the subjective experience (or belief) that the past is fixed and the future malleable. Observing that local recordings of the past (as opposed to radiation from distant stars) are actually available only in the present, we can be sure that the sense of immediate accessibility of a fixed past is an illusion. Indeed, from the point of view of the mathematics 'modification' of the future makes as much sense as modification of the past.

A third source of trouble is the belief in (or sense of) 'free-will', whereby human beings take themselves to be agents acting as a *causeless cause*. From the physical point of view this is nonsense. It is sufficient here to point out that it is a violation of Newton's third law.

These issues illustrate a lack of correspondence between our beliefs and / or subjective experience, and the physics. The power of these illusions is a challenge to our ability to think clearly about issues relating to the symmetry and asymmetry of time. Price [25] in particular has written eloquently on the distorting influence of psychological bias - in particular our sense agency - in discussions of time's arrow.

## Reverse Causation

A few words on the relevance of advanced potentials to the theme of this conference: The use of the phrase 'reverse causation' implies that one can meaningfully (i.e. semantically, if not in practice physically) separate the notion of logical casualty from temporal ordering. In order to do that, one must be able to identify a (more or less universal) property that distinguishes between a cause and an effect that is *not* the temporal order. Some arguments have been given here in support of the rehabilitation of the advanced potential. If one wished to identify all currents as causes and all potentials as effects, then absorption of radiation is an example of reverse causation. Since the most mathematically efficient description of absorption is through (exclusively) advanced potentials (Eq. (8) with $A_{out} = 0$), one may choose to associate reverse causation with the predominance of advanced potentials in an appropriately defined maximally efficient description. But no connection with the flow of entropy has been established in this document. As a result of considerations in the section 'The Cosmological Boundary Condition', it is not clear that entropy necessarily increases in Cosmological time, even in the event that retarded potentials turn out to be predominant in the 'most efficient' description of EM processes.

---

[10] In private discussions Avshalom Elitzur suggested that the common subjective view is a consequence of a 'half-block universe' wherein the moving now is the boundary between a fixed past and an undecided future – grounded somehow in the physics This seems to be the most straightforward projection of our subjective experience onto physics, and deserving therefore of consideration - especially when articulated by a leading researcher. The position adopted in this document however is that investigation of this possibility is best postponed until after one is quite sure that all (unphysical) anthropocentric bias has been thoroughly exposed and expunged from discussion of the physics.

## ACKNOWLEDGMENTS

The author is very grateful to Daniel Sheehan for putting together a remarkably enjoyable conference on such an interesting topic. I learned a lot from the other speakers – thank you to them also. I am grateful to Yurij Baryshev for first alerting me to the need for a deeper understanding of energy non-conservation associated with cosmological red-shift.

## REFERENCES

1. K. A. Milton, *The Casimir Effect*, World Scientific, River Edge, New Jersey, 2001.
2. C. Itzykson and J.-B. Zuber, *Quantum field theory*, McGraw-Hill, New York, 1985.
3. A. D. Fokker, *Zeitschrift für Physik* **58**, 386-393 (1929).
4. K. Schwarzschild, *Gottinger Nachrichten* **128**, 132- (1903).
5. H. Tetrode, *Zeitschrift für Physik* **10**, 317-328 (1922).
6. D. Leiter, "On a new, finite, "charge-field" formulation of classical electrodynamics," in *Foundations of Radiation Theory and Quantum Electrodynamics*, edited by A. O. Barut, Dover, New York, 1980, pp. 195-201.
7. R. P. Feynman, *Physical Review* **76**, 749-759 (1949).
8. M. Ibison, *Fizika A* **12**, 55-74 (2004).
9. P. C. W. Davies, *Journal of Physics A* **4**, 836-845 (1971).
10. P. C. W. Davies, *Journal of Physics A* **5**, 1024-1036 (1972).
11. F. Hoyle and J. V. Narlikar, *Annals of Physics* **54**, 207-239 (1969).
12. F. Hoyle and J. V. Narlikar, *Annals of Physics* **62**, 44-97 (1971).
13. M. Ibison, *Annals of Physics* **321**, 261-305 (2006).
14. J. A. Wheeler and R. P. Feynman, *Reviews of Modern Physics* **17**, 157-181 (1945).
15. J. A. Wheeler and R. P. Feynman, *Reviews of Modern Physics* **21**, 425-433 (1949).
16. P. C. W. Davies, *The Physics of Time Asymmetry*, University of California Press, Berkeley, CA, 1977.
17. P. C. W. Davies, *Journal of Physics A* **5**, 1722-1737 (1972).
18. D. T. Pegg, *Reports on Progress in Physics* **38**, 1339-1383 (1975).
19. J. A. Peacock, *Cosmological Physics*, Cambridge University Press, Cambridge, UK, 1999.
20. Y. V. Baryshev, "Conceptual Problems of the Standard Cosmological Model" in *1st Crisis in Cosmology Conference, CCC-I*, edited by E. J. Lerner and J. B. Almeida, AIP Conference Proceedings 822, Melville, New York, 2006, pp. 23-33.
21. C. W. Misner, K. S. Thorne, and J. A. Wheeler, *Gravitation*, W. H. Freeman and Co., New York, 1973.
22. L. D. Landau and E. M. Lifshitz, *The Classical Theory of Fields*, Pergamon Press, Oxford, UK, 1980.
23. J. D. Jackson, *Classical Electrodynamics*, John Wiley & Sons, Inc., New York, 1998.
24. E. J. Squires, *To Acknowledge the Wonder*: *The Story of Particle Physics*, Adam Hilger, 1985.
25. H. Price, *Time's Arrow and Archimedes' Point*: *New Directions for the Physics of Time*, Oxford University Press, USA, 2006.